\documentclass[aps,pra,twocolumn,superscriptaddress,english]{revtex4-1}
\usepackage{amssymb}
\usepackage{graphicx}
\usepackage{amsmath}
\usepackage{amsthm}
\usepackage{amsfonts}
\usepackage{txfontsb}
\usepackage{color}
\usepackage{bm}
\usepackage{bbm}
\usepackage{url}
\usepackage[driverfallback=dvipdfm]{hyperref}
\hypersetup{pdfpagemode=FullScreen,colorlinks=true,breaklinks,urlcolor=blue,linkcolor=blue,citecolor=blue}

\makeatletter
\def\iddots{\mathinner{\mkern1mu\raise\p@
    \hbox{.}\mkern2mu\raise4\p@\hbox{.}\mkern2mu
    \raise7\p@\vbox{\kern7\p@\hbox{.}}\mkern1mu}}
\def\adots{\mathinner{\mkern2mu\raise\p@\hbox{.} 
 \mkern2mu\raise4\p@\hbox{.}\mkern1mu
 \raise7\p@\vbox{\kern7\p@\hbox{.}}\mkern1mu}}
\makeatother

\begin{document}

\global\long\def\id{\mathbbm{1}}
\global\long\def\ui{\mathbbm{i}}
\global\long\def\ud{\mathrm{d}}

\title{Unified theory to characterize Floquet topological phases by quench dynamics}

\author{Long Zhang}
\affiliation{International Center for Quantum Materials, School of Physics, Peking University, Beijing 100871, China}
\affiliation{Collaborative Innovation Center of Quantum Matter, Beijing 100871, China}

\author{Lin Zhang}
\affiliation{International Center for Quantum Materials, School of Physics, Peking University, Beijing 100871, China}
\affiliation{Collaborative Innovation Center of Quantum Matter, Beijing 100871, China}

\author{Xiong-Jun Liu}
\thanks{Correspondence addressed to: xiongjunliu@pku.edu.cn}
\affiliation{International Center for Quantum Materials, School of Physics, Peking University, Beijing 100871, China}
\affiliation{Collaborative Innovation Center of Quantum Matter, Beijing 100871, China}
\affiliation{CAS Center for Excellence in Topological Quantum Computation, University of Chinese Academy of Sciences, Beijing 100190, China}
\affiliation{Institute for Quantum Science and Engineering and Department of Physics, Southern University of Science and Technology, Shenzhen 518055, China}
\affiliation{Synergetic Innovation Center for Quantum Effects and Applications, Hunan Normal University, Changsha 410081, China}


\begin{abstract}
The conventional characterization of periodically driven systems usually necessitates the time-domain information beyond Floquet bands,
hence lacking universal and direct schemes of measuring Floquet topological invariants. Here we propose a unified theory based on quantum quenches
to characterize generic $d$-dimensional ($d$D) Floquet topological phases, in which the topological invariants are constructed
with only minimal information of the static Floquet bands. For a $d$D phase which is initially static and trivial,
we introduce the quench dynamics by suddenly turning on the periodic driving, and show that the quench dynamics
exhibits emergent topological patterns in ($d-1$)D momentum subspaces where Floquet bands cross,
from which the Floquet topological invariants are directly obtained.
This prediction provides a simple and unified characterization, in which one can not only extract the number of
conventional and anomalous Floquet boundary modes, but also identify the topologically protected singularities in the phase bands.
The applications are illustrated with 1D and 2D models which are readily accessible in cold atom experiments.
Our study opens a new framework for the characterization of Floquet topological phases.
\end{abstract}

\maketitle

{\em Introduction.---}Topological phases have been extensively explored in equilibrium systems~\cite{TI_review1,TI_review2}.
Recently, Floquet topological phases were proposed in periodically driven systems and attract broad interests~\cite{Oka2009,Kitagawa2010,Lindner2011,Cayssol2013,Rudner2019}.
The spatiotemporal control introduces an additional degree of freedom in the time domain and leads to richer Floquet topological phases~\cite{Rudner2013,Nathan2015,Carpentier2015,Roy2017,Yao2017}.
As a versatile tool for quantum engineering~\cite{Goldman2014,Bukov2015,Eckardt_Review}, the
periodic driving opens a new route to realize various topological states~\cite{Rechtsman2013,Jotzu2014,Zheng2014,Usaj2014,Flaschner2016,Rodriguez2019,Bomantara2019,Peng2019,Seshadri2019,Yang2019,Hu2020}.

Like the Bloch momentum states in a spatial lattice, quasienergy states with the Floquet bands are obtained in a periodically driven system,
and can be characterized by the time-independent effective Floquet Hamiltonian~\cite{Eckardt_Review}.
The inclusion of the temporal degree of freedom brings about new classes of topological phases that cannot be completely interpreted by the topological invariants of the static Floquet bands.
Such phases are known as the ``anomalous" Floquet topological phases~\cite{Rudner2013,Hu2015,Mukherjee2017,Maczewsky2017,Wintersperger2020},
where the bulk topology has no direct correspondence to the number of boundary modes~\cite{Rudner2013}.
Their characterization was proposed by considering the micromotion throughout the entire driving period~\cite{Rudner2013,Nathan2015,Fruchart2016,Unal2019},
with the topology being classified either by the winding number defined in the momentum-time space~\cite{Rudner2013,Fruchart2016},
or by topologically protected singularities in the spectrum of the time-evolution operator called the phase bands~\cite{Nathan2015}.
Both of them are however physically not intuitive and lack direct measurements in general.

In this Letter, we propose a unified theory with highly feasible characterization scheme, in which
the Floquet topological invariants are extracted based on minimal information of the Floquet bands.
We consider a class of $d$-dimensional ($d$D) models of Floquet topological phases,
with periodic driving applied on top of the $d$D band structures. Instead of studying the steady state, 
we develop the characterization theory using the quench dynamics from an initially fully-polarized trivial state,
which is induced by suddenly turning on the periodic driving.
Our theory is built by generalizing a fundamental dynamical bulk-surface correspondence~\cite{Zhanglin2018,Zhanglong2019a,Zhanglong2019b,Yu2020},
which shows that the bulk topology of a $d$D equilibrium phase has a one-to-one correspondence to quench-induced
dynamical topological patterns emerging on ($d-1$)D momentum subspaces called band inversion surfaces (BISs).
This result has been experimentally observed in nondriven cold-atom~\cite{Sun2018,Song2019,Yi2019,Wang2020}
and solid-state spin~\cite{WangY2019,Niu2020,Xin2020,Ji2020} systems, applied to non-Hermitian systems~\cite{Zhou2018,Zhou2019,ZhuBo2020},
{ and the dynamical characterization was recently extended to generic quenches from a trivial or nontrivial phase via loop unitary construction~\cite{Hu2020b}.}
Generalizing the dynamical bulk-surface correspondence to Floquet systems as proposed here provides a unified theory for a full characterization of Floquet topological phases.

Our main results include: (i) For $d$D systems with {generic} periodic driving {$V(t)=V(t+T)$,} 
the Floquet topological phases are fully characterized by the $(d-1)$D emergent topology of quench dynamics, induced by suddenly turning on the driving at $t=0$ and measured through stroboscopic time-averaged spin textures, on both static and driving-induced BISs.
(ii) {If tuning an initial phase $\phi$ in the driving $V(t-t_\phi)$ with $t_\phi=(\phi/2\pi)T$,}
the dynamical characterization exhibits distinct features in odd and even dimensional phases.
For odd dimensions, the phases necessitates symmetry-protection, {requiring that $V(t_\phi+t)=V(t_\phi-t)$}, and the characterization scheme is applied after a modification
that dynamical spin textures are averaged from a modified initial time $t_\phi$.
For even dimensions, the phases necessitate no symmetry-protection and the dynamical characterization is generally valid.
(iii) Both the conventional and anomalous Floquet topological phases are characterized by the present unified theory.
Moreover, when tuning $\phi$, the dynamical spin textures in even dimensions also exhibit nontrivial topology in the $\phi$-parameter space,
which corresponds to topological singularities in the time-dependent phase bands.
(iv) The Floquet topological invariants are directly measurable in this theory, and experimental models are proposed.

\begin{figure}
\includegraphics[width=0.47\textwidth]{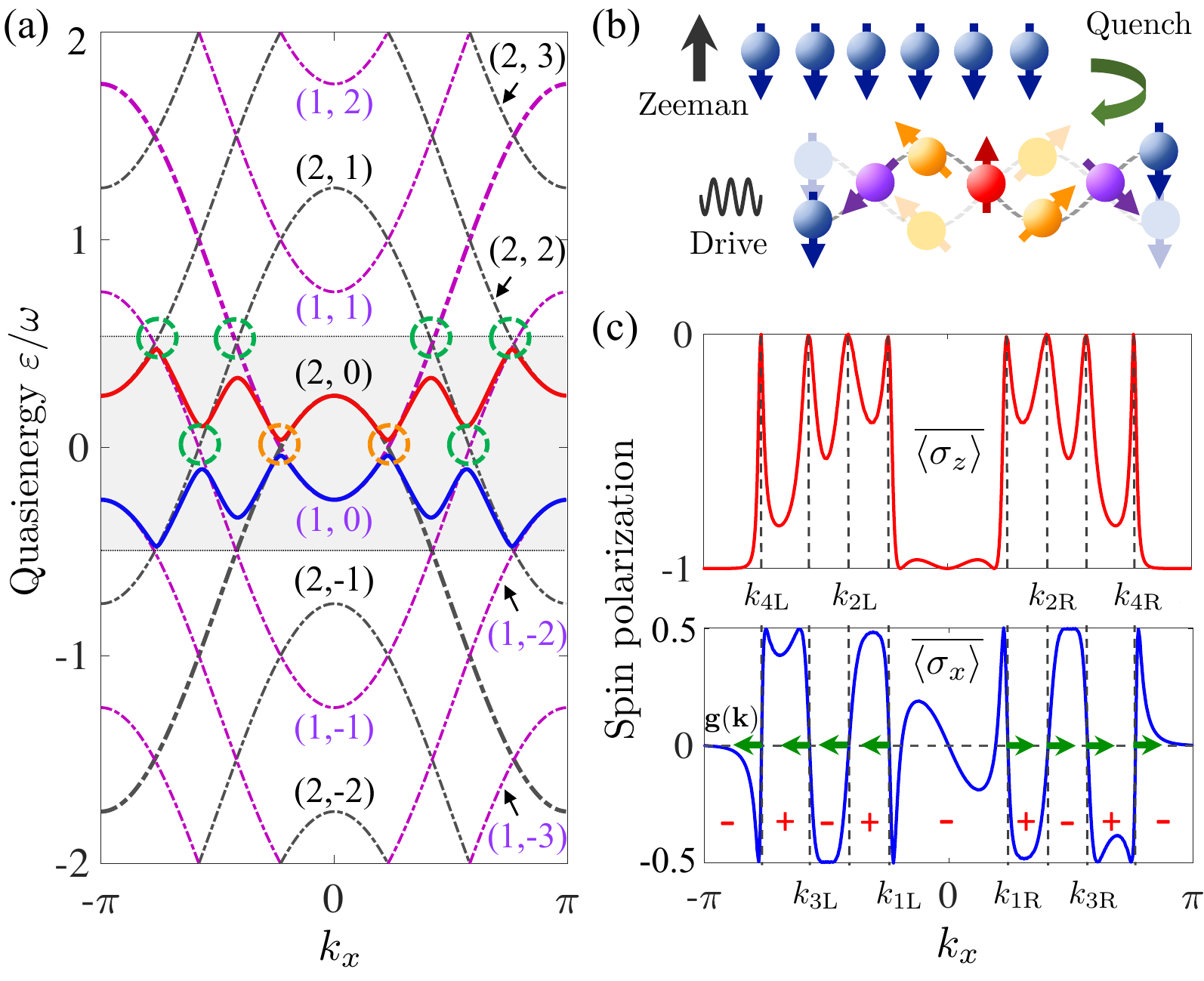}
\caption{Characterization scheme for the 1D model.
(a) Band structure.
Copies of spin-up and -down bands (dot-dashed curves) are labeled by $(s,m)$, which define both driving-induced (dashed green circles) and static BISs (dashed orange circles).
The Floquet bands (solid curves) are shown in the FBZ $[-\pi/T,\pi/T]$ (shaded region). 
(b) Quench protocol.  When $t<0$, a large Zeeman field is applied; at $t=0$, the Zeeman field is removed and periodic driving is turned on.
(c) Stroboscopic time-averaged spin textures $\overline{\langle\sigma_z\rangle}$ (upper) and $\overline{\langle\sigma_x\rangle}$ (lower).
The vanishing spin polarizations determine the BISs, each corresponding to a pair of momenta $k_x=k_{j{L/R}}$ ($j=1,2,3,4$).
The dynamical spin-texture field ${\bm g}({\bm k})$ on BISs
(green arrows) characterizes the topology. The sign ``+'' (``-'') denotes the region where $h_{F,z}>0$ ($<0$).
Here $t_{\rm so}=0.3t_0$, $m_z=1.5t_0$, $\omega=2t_0$ and $V_0=t_0$.
}\label{fig1}
\end{figure}

{\em Generalized bulk-surface duality}.--To illustrate how the dynamical characterization scheme works, we start with a simple 1D model with harmonic driving described by
\begin{align}\label{Ham_t}
H(t)=H_{\rm s}+Ve^{\ui\omega t}+V e^{-\ui\omega t}.
\end{align}
Here $H_{\rm s}={\bm h}\cdot{\bm\sigma}=(m_z-2t_0\cos k_x)\sigma_z+2t_{\rm so}\sin k_x\sigma_x$ can be realized in a 1D optical Raman lattice~\cite{LiuXJ2013,Song2018},
with $\sigma_{x,y,z}$ being the Pauli matrices, and $t_0$ ($t_{\rm so}$) denoting the spin-conserved (-flipped) hopping.
The driving $V=V_0\sigma_z$ can be achieved by modulating bias magnetic field.

We characterize the bulk topology via BISs where the spin-up and spin-down bands cross.
For the static Hamiltonian $H_s$, the BIS, existing if $|m_z|<2t_0$,
corresponds to two momenta with $h_z(k_x)=0$~\cite{Zhanglin2018},
and the spin-orbit (SO) coupling $h_x$ opens a topological gap at the BIS.
The quasi-energy spectra of the total Hamiltonian $H(t)=H(t+T)$ ($T=2\pi/\omega$)
are given by the quasi-energy operator $Q(t)\equiv{H}(t)-\ui\partial_t$~\cite{Eckardt2015,Supp}.
The spin-up and spin-down bands are copied and shifted by energies $m\omega$,
leading to more band crossings induced by periodic driving, as shown in Fig.~\ref{fig1}(a).
These bands are labeled by $(s, m)$ in sequence, where $s=1$ ($2$) denotes the spin-up (-down) and $m$ represents the shifted energy $m\omega$.
In particular, the band $(1,0)$ has crossings with three bands $(2,1)$, $(2,2)$ and $(2,3)$ for $\omega=2t_0$.
The finite $V$ and SO terms further open gaps at these BISs,
rendering the Floquet bands characterized by the effective Hamiltonian $H_F\equiv\ui\log U(T)/T$,
with $U(t)={\cal T}\exp\big[-\ui\int_{0}^t H(\tau)d\tau\big]$ and ${\cal T}$ denoting the time ordering.
The Floquet bands in the Floquet Brillouin zone (FBZ) $[-\pi/T,\pi/T]$ are shown in Fig.~\ref{fig1}(a) (solid curves in the shaded region).

We generalize the bulk-surface duality, which reduces the $d$D bulk topology to a $(d-1)$D topological invariant defined on BISs~\cite{Zhanglin2018},
to Floquet systems. First, the present 1D Floquet Hamiltonian $H_F=h_{F,x}\sigma_x+h_{F,z}\sigma_z$ maintains the chiral symmetry $S=\sigma_y$ of $H_s$. For the $j$th BIS with a pair of momenta $k_x=k_{j{L,R}}$ satisfying $h_{F,z}(k_x)=0$, the invariant reads
$\nu_j=\big({\rm sgn}[h_{F,x}(k_{j{R}})]-{\rm sgn}[h_{F,x}(k_{j{L}})]\big)/2$.
Further, we divide the BISs, each formed by two bands $(1, m_1)$ and $(2, m_2)$, into two categories by the integer $\delta m\equiv|m_1-m_2|$:
If $\delta m$ is odd (even), BISs are formed at the border (center) of FBZ and denoted as $\pi$-BISs (0-BISs)~\cite{Supp}.
The classification of BISs is periodic in quasienergy.
We obtain the topological number of the lower Floquet band in FBZ by
\begin{align}\label{W_nu}
W=&\sum_j (-1)^{q_j/\pi}\nu_{j},
\end{align}
where $q_j=0$ or $\pi$ for the $q_j$-BIS.
Finally, the invariants $W_{0/\pi}=\sum_{q_j=0/\pi}\nu_{j}$ give the numbers of boundary modes in the two different gaps~\cite{Supp}.

\begin{figure}
\includegraphics[width=0.45\textwidth]{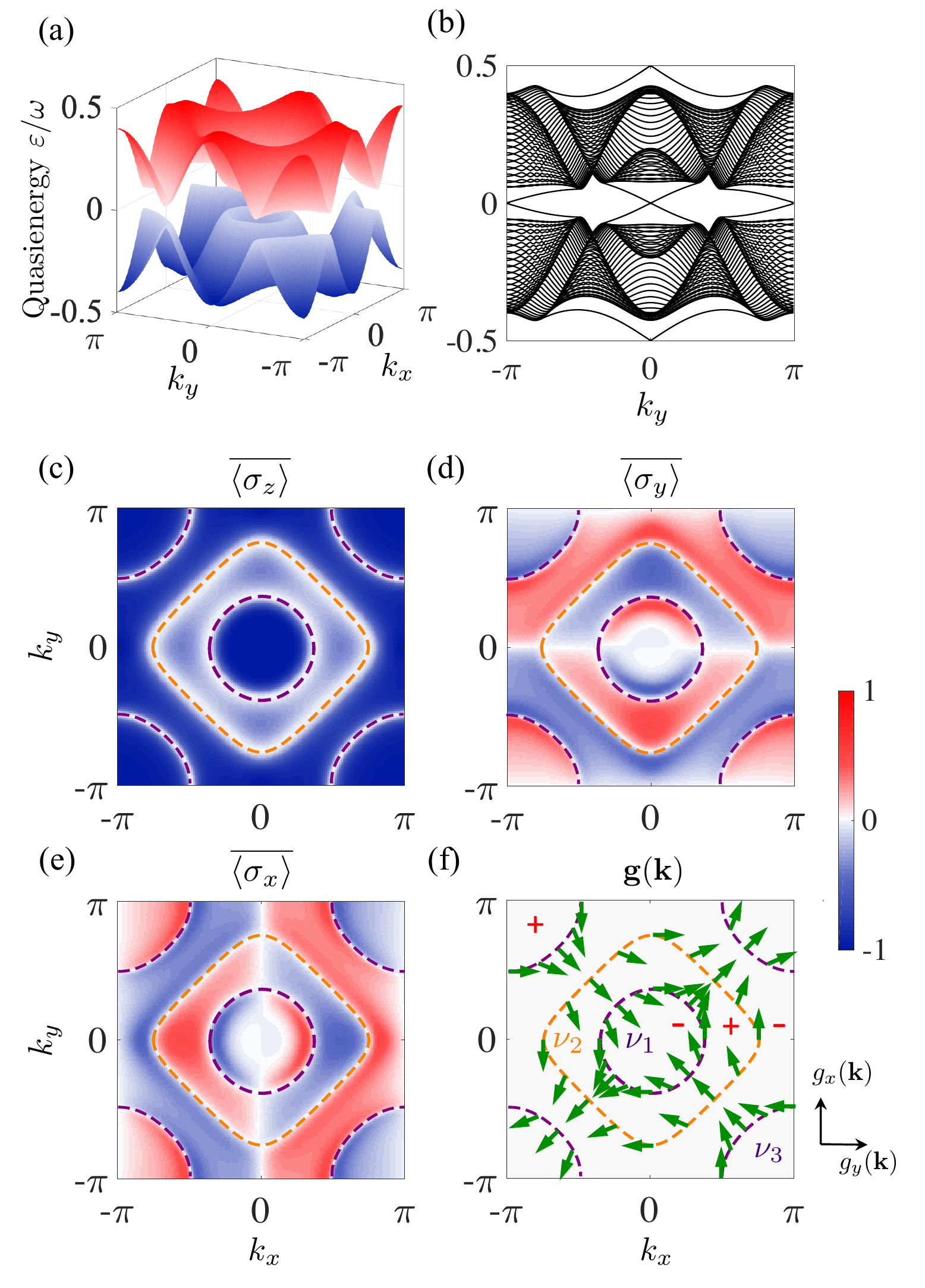}
\caption{Dynamical characterization of the 2D model. (a-b) Quasienergy spectrum in the FBZ
with periodic boundary conditions (a) or for a cylindrical geometry (b).
(c-e) Stroboscopic time-averaged spin textures $\overline{\langle\sigma_{i}({\bm k})\rangle}$ ($i=x,y,z$).
Three rings emerge in all the textures (dashed curves):
One (orange) is the $\pi$-BIS and two (purple) are 0-BISs according to the band structure in (a).
The inner one surrounding the $\Gamma$ point corresponds to the static BIS.
(f) Dynamical spin-texture field ${\bm g}({\bm k})$ (green arrows) is constructed,
whose winding characterizes the topological invariants {$\nu_{1,2}=-1$ and $\nu_3=+1$}.
The sign ``+'' (``-'') denotes the region where $h_{F,z}>0$ ($<0$).
Here $t_{\rm so}=0.5t_0$, $m_z=3t_0$, $\omega=5t_0$ and $V_0=2t_0$.
}\label{fig2}
\end{figure}

{\em Dynamical characterization.--}The quantum dynamics is induced by suddenly quenching
an initially fully polarized trivial phase with $m_z\gg t_0,\omega$ to Floquet topological regime described by Eq.~\eqref{Ham_t} at time $t=0$ [Fig.~\ref{fig1}(b)].
The BISs and bulk-surface duality can be characterized dynamically as in Refs.~\cite{Zhanglin2018,Zhanglong2019a,Zhanglong2019b},
but now based on the stroboscopic time-averaged spin textures~\cite{note0}
\begin{align}
\overline{\langle\sigma_i({\bm k})\rangle}=\lim_{N\to\infty}\frac{1}{N}\sum_{n=0}^N \langle\sigma_i({\bm k},t=nT)\rangle, \ i=x,y,z.
\end{align}
{Here $\langle\sigma_i({\bm k},t)\rangle={\rm Tr}\big[\rho_0({\bm k}) U^\dagger({\bm k},t)\sigma_iU({\bm k},t)\big]$,
where $\rho_0$ is the density matrix of the initial state.}
The spin textures $\overline{\langle\sigma_{x,z}(k_x)\rangle}$ are shown in Fig.~\ref{fig1}(c).
There are four pairs of momenta at which $\overline{\langle\sigma_{x,z}(k_x)\rangle}=0$, indicating the momenta at four BISs $k_x=k_{jL/R}$~\cite{note1}.
The topological invariant on each BIS is further characterized by a dynamical spin-texture field ${\bm g}({\bm k})$, which has only one component here $g_x(k_x)=-\partial_{k_\perp}\overline{\langle\sigma_x(k_x)\rangle}/{\cal N}_k$ and equals $h_{F,x}/|h_{F,x}|$ on BISs, with
${\cal N}_k$ being the normalization factor and $k_\perp$ pointing from the region $h_{F,z}({\bm k})<0$ to $h_{F,z}({\bm k})>0$.
The dynamical field characterizes the topological invariant $v_j=+1$ on each BIS with the parameters given in Fig.~\ref{fig1},
so $W=0$, but $W_0=W_\pi=2$, giving the number of edge states within each gap,
and rendering an anomalous Floquet topological phase~\cite{Supp}.

For even dimensions, we take a 2D model as an example to illustrate the dynamical characterization scheme, with the Hamiltonian taking
the similar form (\ref{Ham_t}). The static Hamiltonian is now chosen as the quantum anomalous Hall model $H_{\rm s}({\bm k})={\bm h}({\bm k})\cdot{\bm\sigma}$ with ${\bm h}({\bm k})=(t_{\rm so}\sin k_x,t_{\rm so}\sin k_y,m_z-t_0\cos k_x-t_0\cos k_y)$~\cite{LiuXJ2014}, which has been realized in cold atoms~\cite{Wu2016,Sun2017}.
The Floquet bands are shown in Fig.~\ref{fig2}(a) with periodic boundary conditions, and also in Fig.~\ref{fig2}(b) for a cylindrical
geometry, with periodic boundary conditions only in the $y$ direction.
The stroboscopic time-averaged spin textures $\overline{\langle\sigma_{i}({\bm k})\rangle}$ ($i=x,y,z$) are shown in Fig.~\ref{fig2}(c-e).
One finds that three ring-shaped structures emerge in all the textures, identified as two 0-BISs and one $\pi$-BIS according to the band structure.
The dynamical field ${\bm g}({\bm k})=(g_y,g_x)$ is obtained by $g_i({\bm k})=-\partial_{k_\perp}\overline{\langle\sigma_i({\bm k})\rangle}/{\cal N}_k$.
Along each ring, ${\bm g}({\bm k})$ winds only once [see Fig.~\ref{fig2}(f)], characterizing the invariants {$\nu_1=-\nu_3=-1$} at $0$-BISs and $\nu_2=-1$ at the $\pi$-BIS. 
The Chern number of lower subband reads ${\rm Ch}=+1$ and topological invariants {$(W_0,W_\pi)=(0,-1)$} indicate the existence of { anomalous edge states~\cite{note}.}

{\em Generic $d$-dimensional systems.---}We now turn to the {generic theory} for $d$D driven systems described by {
\begin{align}\label{GFHam}
\begin{split}
H({\bm k},t)=&{H}_{\rm s}({\bm k})+V({\bm k},t),\quad H_{\rm s}({\bm k})=\sum_{i=0}^d h_i({\bm k})\gamma_i,\\
V({\bm k},t)=&V_{l_1}({\bm k},t)\gamma_{l_1}+V_{l_2}({\bm k},t)\gamma_{l_2}+\cdots,
\end{split}
\end{align}
where generically $V_{l_i}(t)=V_{l_i}(t+T)$ with $l_i\in\{0,1,\cdots,d\}$.} If $d$ is {odd,
a constraint $V(t_\phi+t)=V(t_\phi-t)$ ($0\leq t_\phi<T$) is necessary to ensure a chiral symmetry~\cite{Supp}.}
The $\gamma$ matrices satisfy $\{\gamma_i,\gamma_j\}= 2\delta_{ij}$, and are
of dimensionality $n_d = 2^{d/2}$ (or $2^{(d+1)/2}$) if
$d$ is even (or odd).
{One sees that $t_\phi=0$ in the above 1D model (\ref{Ham_t}).}

In odd dimensions, the static Hamiltonian $H_{\rm s}({\bm k})$ necessitates a chiral symmetry $S=\ui^{(d+1)/2}\prod_{i=0}^d\gamma_i$. 
When {$t_\phi=0$}, $S H({\bm k},t)S^{-1}=-H({\bm k},-t)$ and $S U({\bm k},t)S^{-1}=U({\bm k},-t)$, which leads to
$S H_F({\bm k})S^{-1}=-H_F({\bm k})$.
Therefore, for any $d$, even or odd {(with $t_\phi=0$)}, we have $H_F({\bm k})=h_{F,0}({\bm k})\gamma_0+\sum_{i=1}^d h_{F,i}({\bm k})\gamma_i$.
Note that the $\gamma$ matrices here are those in $H_{\rm s}$.
Similar to the convention in Ref.~\cite{Zhanglin2018}, we choose
$h_{F,0}({\bm k})$ to characterize the dispersion of $n_d$ decoupled bands, and define the BISs by $h_{F,0}({\bm k})=0$.
The remaining components are denoted as the SO field ${\bm h}^F_{\rm so}({\bm k})\equiv(h_{F,1},h_{F,2},\cdots,h_{F,d})$.
For dynamical characterization, we choose an initial state that is fully polarized in the $\gamma_0$ axis,
which leads to ${\rm BIS}=\{{\bm k}\big\vert\overline{\langle\gamma_i({\bm k})\rangle}=0, \forall i\}$
and ${\bm g}({\bm k})=(g_1,g_2,\cdots,g_d)$ with $g_i({\bm k})=-\partial_{k_\perp}\overline{\langle\gamma_i({\bm k})\rangle}/{\cal N}_k$.
Here $k_\perp$ points from the region $h_{F,0}({\bm k})<0$ to $h_{F,0}({\bm k})>0$.
One can check that ${\bm g}({\bm k})\simeq {\bm h}^F_{\rm so}({\bm k})$ on BISs~\cite{Supp}.
The topological invariant $\nu_j$ is then characterized by the winding of ${\bm g}({\bm k})$ on the $j$-th BIS,
and $W=W_0-W_\pi$, where
\begin{align}
W_{q}=\sum_{j\in q\textrm{-BIS}}\frac{\Gamma(\frac{d}{2})}{2\pi^{\frac{d}{2}}(d-1)!}\int_{{\rm BIS}_{j}}{\bm g}(\ud {\bm g})^{d-1},\ q=0,\pi.
\end{align}
Here $\Gamma(x)$ is the Gamma function, ${\bm g}(\ud{\bm g})^{d-1}\equiv\epsilon^{i_{1}i_{2}\cdots i_{d}}g_{i_{1}}\ud g_{i_{2}}\wedge\cdots\wedge\ud g_{i_{d}}$
with $\epsilon^{i_{1}i_{2}\cdots i_{d}}$ being the fully anti-symmetric tensor and $i_{1,2,\dots,d}\in\{1,2,\dots,d\}$,
and `$\ud$' denotes the exterior derivative.
Compared to other characterization methods~\cite{Rudner2013,Nathan2015,Unal2019},
the above invariants are characterized in only {\it lower}-dimensional momentum space and are easily measurable in experiments.

\begin{figure}
\includegraphics[width=0.49\textwidth]{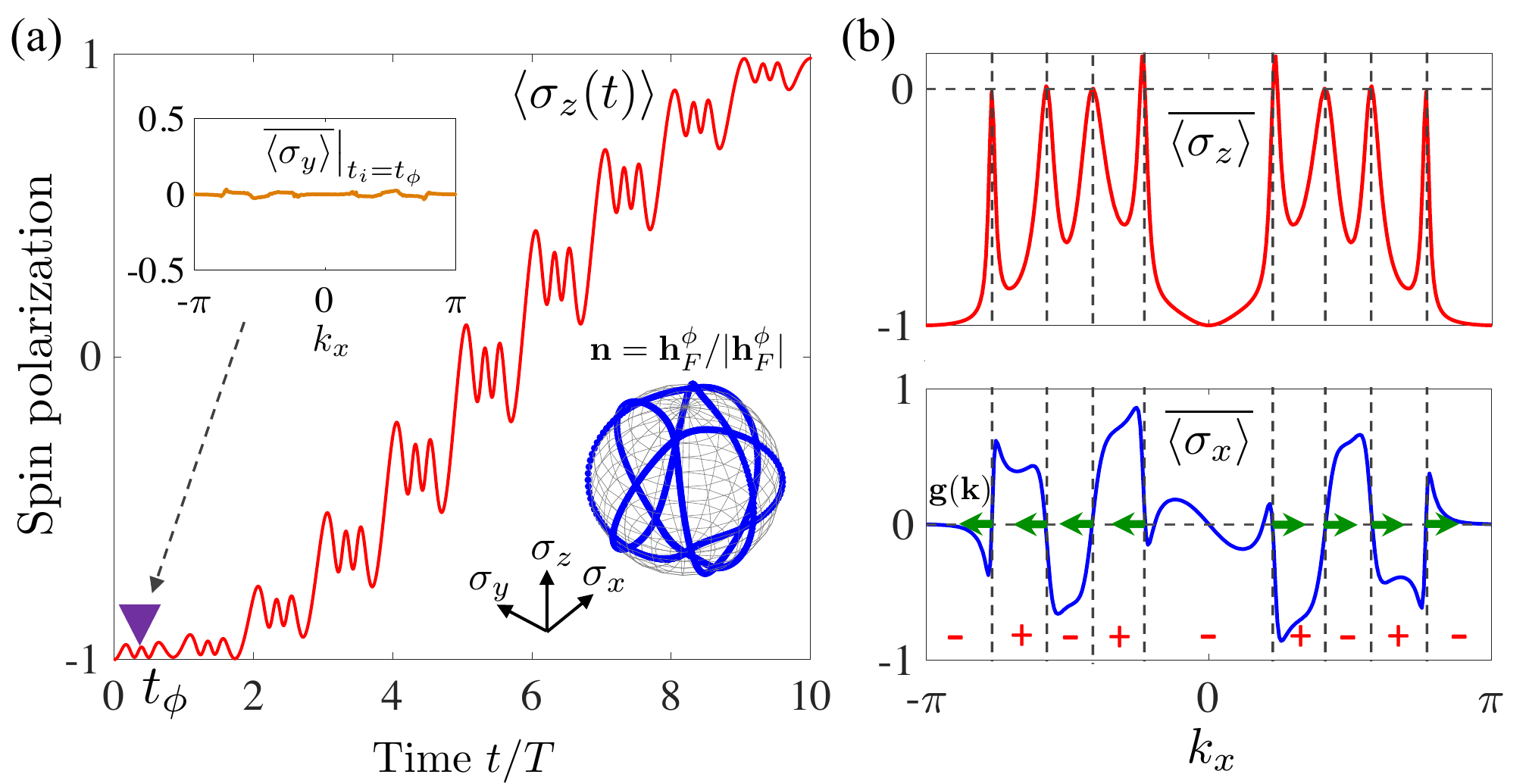}
\caption{Dynamical characterization of the 1D model with $\phi=3\pi/4$.
(a) The stroboscopic time average of spin dynamics should be taken from a modified initial time $t_\phi$
such that $\overline{\langle\sigma_y\rangle}$ is almost zero for all $k_x$
(the upper-left inset). The lower right inset: The unit vector field of $H_F^\phi$ has a nonzero component in the $\sigma_y$ axis.
(b) Dynamical characterization with the initial time $t_\phi$.
The results are consistent with those in Fig.~\ref{fig1}(c).
Here the parameters are taken the same as in Fig.~\ref{fig1}.
}\label{fig3}
\end{figure}

\begin{figure}
\includegraphics[width=0.45\textwidth]{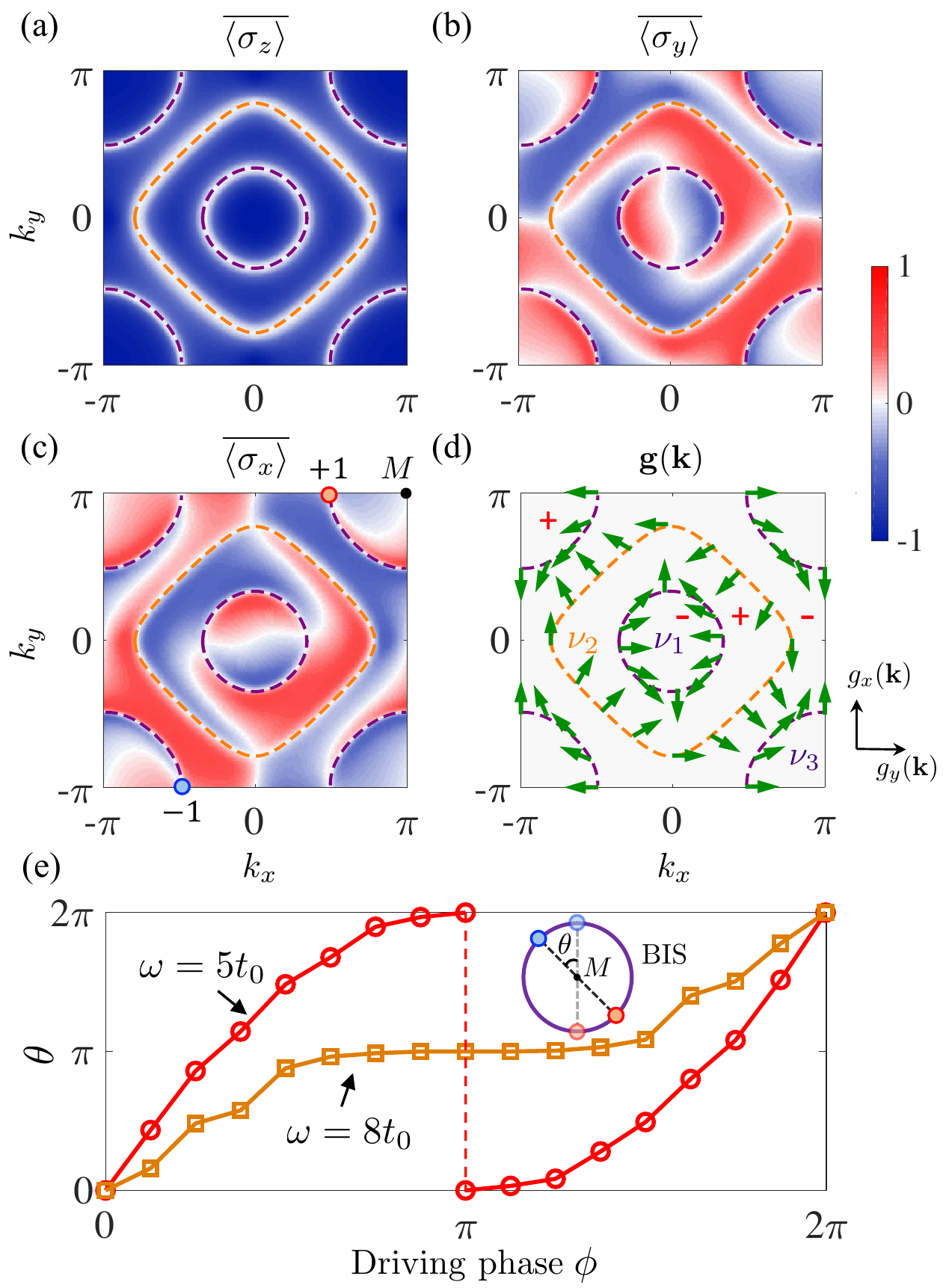}
\caption{Dynamical characterization of the 2D model with $\phi\neq0$.
(a-d) Stroboscopic time-averaged spin textures $\overline{\langle\sigma_{i}({\bm k})\rangle}$ (a-c) and the constructed
dynamical spin-texture field ${\bm g}({\bm k})$ (d) for $\phi=\pi/2$. Other parameters are taken the same as in Fig.~\ref{fig2}.
(e) The rotation angle $\theta$ of the two poles $h^\phi_{F,y}/|{\bm h}^\phi_{F}|=\pm1$ on the BIS surrounding the $M$ point [see (c)] versus the driving phase $\phi$.
When $\omega=5t_0$ (a-c), the poles rotate along the BIS twice with the increasing of $\phi$.
As a comparison, another case with $\omega=8t_0$ is also plotted, where the two poles rotate around only once. 
Details can be found in Ref.~\cite{Supp}.
}\label{fig4}
\end{figure}

{\em {Tuning the} driving phase.---}Now we consider {a nonzero $t_\phi\equiv(\phi/2\pi)T$, with the phase $\phi$} controlled by modulating bias magnetic field. 
Denote by $H_F^{\phi}$ the Floquet Hamiltonian and write $H_F\equiv H_F^{\phi=0}$ for simplicity.
We have the relation $H_F^{\phi}=U^\dagger(t_\phi)H_FU(t_\phi)$~\cite{Supp}.
Since $U(t_\phi)$ is local unitary, $H_F^{\phi}$ and $H_F$ have the same topology~\cite{Chen2010}.

In odd dimensions, 
the Floquet Hamiltonian $H_F^{\phi}$ has a different chiral symmetry $S_{\phi}=U^\dagger(t_\phi)S U(t_\phi)$, which indicates that $H_F^{\phi}$ cannot be written in terms of the same $\gamma$ matrices as $H_F$.
For this instead of performing a direct measurement of $\overline{\langle\gamma_i({\bm k})\rangle}$,
the stroboscopic time average is now taken from a modified initial time $t_\phi$ as
\begin{align}~\label{gammai_phi}
\overline{\langle\gamma_i({\bm k})\rangle}=\lim_{N\to\infty}\frac{1}{N}\sum_{n=0}^N \langle\gamma_i({\bm k},t_\phi+nT)\rangle.
\end{align}
This is equivalent to quenching an initial state with incomplete polarization and characterizing the topology of $H_F$ (rather than $H_F^\phi$)~\cite{Supp}.
We show the numerical results in Fig.~\ref{fig3} for the 1D model with {harmonic driving $V(t)=V_0\cos(\omega t-\phi)\sigma_z$ and $\phi=3\pi/4$. The numerical confirmation with square-wave driving, which is polychromatic, is given in Supplementary Material~\cite{Supp}.}
One finds that $H_F^\phi$ has a nonzero component $h_{F,y}^\phi$ which breaks $\sigma_y$-chiral symmetry [the lower-right inset in Fig.~\ref{fig3}(a)].
However, by choosing the initial time $t_\phi$, which gives $\overline{\langle\sigma_y({\bm k})\rangle}=0$ in quench dynamics (the upper-left inset), the Floquet topological phase is again well characterized [Fig.~\ref{fig3}(b)].

More nontrivial results are obtained in even dimensions, where the Floquet topological phases necessitate no symmetry-protection and the dynamical scheme is directly applicable.
We write $H_F^{\phi}={\bm h}_F^\phi({\bm k})\cdot{\bm\gamma}=\sum_{i=0}^d h^\phi_{F,i}({\bm k})\gamma_i$.
The BISs and topological patterns are defined with respect to ${\bm h}^\phi_{F}({\bm k})$.
Here we still take the above 2D model as an example, but set a nonzero phase $\phi$. The results are shown in Fig.~\ref{fig4}(a-d) for $\phi=\pi/2$. One can see
that although time-averaged spin textures are different from those in Fig.~\ref{fig2},
the two characterizations yield the same classification of the topological phase.

The key finding is that by varying $\phi$, the rotation of ${\bm h}^\phi_{F}({\bm k})$ can identify the emergence of {\it driving}-induced ``topological singularities'' in the phase bands of $U(t_\phi)$~\cite{Nathan2015}.
We consider ${\bm h}^\phi_{F}({\bm k})$ on the BIS surrounding the $M$ point,
whose distribution is signalled by the location of the two poles where $h^\phi_{F,y}/|{\bm h}^\phi_{F}|=\pm1$
[see Fig.~\ref{fig4}(c)] (or equivalently, choose the poles $h^\phi_{F,x}/|{\bm h}^\phi_{F}|=\pm1$).
We track the rotation of the pole positions measured from $\overline{\langle\sigma_{x}({\bm k})\rangle}$ 
during one driving period $0< t_\phi< T$ [Fig.~\ref{fig4}(e)].
The results show that every full rotation marks the emergence of
a band-touching singularity at the $M$ point~\cite{Supp}, with the number of driving-induced singularities given by
\begin{align}
N_{\rm DS}=\frac{1}{2\pi}\oint_\phi d\theta(\phi),
\end{align}
where $\theta$ denotes the rotation angle [the inset of Fig.~\ref{fig4}(e)].
Each singularity corresponds to a driving-induced BIS and contribute to topological number of Floquet bands.

{Before conclusion we point out that the dynamical characterization theory proposed here is fully accessible based on current experiments~\cite{Supp}.
Firstly, the static part $H_s$ of the Floquet Hamiltonian has been widely studied in experiments for 1D to 3D cases~\cite{Song2018,Wu2016,Sun2017,WangY2019,Niu2020,Xin2020,Ji2020}.
Secondly, the periodic driving term can be readily generated by modulating the bias magnetic field. Finally, the quench study of the generalized dynamical bulk-surface correspondence in the present Floquet systems basically follows that carried out in nondriven systems, as having been already achieved in many experiments~\cite{Sun2018,Song2019,Yi2019,Wang2020,WangY2019,Niu2020,Xin2020,Ji2020}.}

{\em Conclusion.---}
We have proposed a dynamical classification theory with quantum quenches
to fully characterize $d$D Floquet topological phases via only minimal information of the Floquet bands.
The characterization is built on a generalized dynamical bulk-surface correspondence,
which shows that the Floquet topological invariants can be directly obtained from the emergent topology of quench dynamics on two ($0$ and $\pi$) types of BISs. Our dynamical theory provides a unified characterization and is highly feasible in experiment: Based on the concept of BISs,
one can precisely determine the both conventional and anomalous boundary modes, and also identify the emergence of
topological singularities in the phase bands. 
This work shall advance the field of Floquet topological phases in both theory and experiment.

This work was supported by National Natural Science Foundation
of China (Grants No. 11825401, No. 11761161003, and
No. 11921005), the National Key R\&D Program of China
(Project No. 2016YFA0301604), and the Strategic Priority Research Program of
Chinese Academy of Science (Grant No. XDB28000000).


\setcounter{equation}{0} \setcounter{figure}{0}
\setcounter{table}{0} 
\renewcommand{\theparagraph}{\bf}
\renewcommand{\thefigure}{S\arabic{figure}}
\renewcommand{\theequation}{S\arabic{equation}}

\onecolumngrid
\flushbottom
\newpage

\section*{\normalsize SUPPLEMENTAL MATERIAL}

In this Supplemental Material, we provide details of dynamical characterization scheme (Sec.~\ref{Sec1}), the relation with boundary states (Sec.~\ref{Sec2}), 
the detection of topological singularities in phase bands (Sec.~\ref{Sec3}), and experimental realization (Sec.~\ref{Sec4}).

\section{Generic dynamical characterization scheme}~\label{Sec1}

In the following, we introduce our dynamical characterization scheme for generic  $d$-dimensional ($d$D) driven systems described by the Hamiltonian (4) in the main text.
We shall first consider the harmonic driving (Secs.~\ref{Sec1A} and~\ref{Sec1B}), and then generalize the results to general driving fields (Sec.~\ref{Sec1C}).
We emphasize that the $\gamma$ matrices in the static Hamiltonian $H_{\rm s}({\bm k})=\sum_{i=0}^d h_i({\bm k})\gamma_i$ 
are of dimensionality $n_d=2^n$, and set to satisfy the trace property ${\rm Tr}[\prod_{j=0}^d\gamma_j]=(-2\ui)^n$
for even $d=2n$ or ${\rm Tr}[S\prod_{j=0}^d\gamma_j]=(-2\ui)^n$ for odd $d=2n-1$, with $S=\ui^n\prod_{j=0}^d\gamma_j$
being the chiral matrix~\cite{Zhanglin2018_S}. For example, in two dimensions we should have ${\rm Tr}[\gamma_0\gamma_1\gamma_2]=-2\ui$; if $\gamma_0=\sigma_z$,
one should set $\gamma_1=\sigma_y$ and $\gamma_2=\sigma_x$.

\subsection{Generalized bulk-surface duality}~\label{Sec1A}

We first consider $H_{\rm s}$ only, which describes a  $d$D gapped topological phase classified by the winding number (for odd $d$) or the $n$-th Chern number (for even $d=2n$).
For such a system, the bulk-surface duality is applicable,
which claims that the topological characterization can reduce to a
$(d-1)$D invariant $W_{d-1}$ defined in band-crossing subspaces dubbed as band inversion surfaces (BISs)~\cite{Zhanglin2018_S}.
Without loss of generality, one can choose $h_{0}(\mathbf{k})$ to describe the band structure with $\mathrm{BIS}\equiv\{{\bm k}\vert h_{0}({\bm k})=0\}$,
and the topological invariant $W_{d-1}$ then corresponds to the winding of spin-orbit (SO) field ${\bm h}_{\mathrm{so}}({\bm k})\equiv(h_{1},\dots,h_{d})$ on all $(d-1)$D BISs.
Note that $W_{d-1}$ is defined to characterize the topology of the {\it lower} $n_d/2$ bulk energy bands;  
the topology of the upper $n_d/2$ bands is then characterized by $-W_{d-1}$.

We then generalize the bulk-surface duality to periodically driven systems.
For a driven system with $H(t)=H(t+T)$, one can employ the Floquet theorem, and use the Floquet states $|\psi_n(t)\rangle=e^{-\ui \varepsilon_n t}|u_n(t)\rangle$,
where $|u_n(t)\rangle=|u_n(t+T)\rangle$ and $\varepsilon_n$ are the quasienergies (see, e.g., Ref.~\cite{Eckardt_review_S} and references therein).
The time-dependent Schr\"odinger equation then reduces to the eigenvalue equation
$Q(t)|u_{n}(t)\rangle=\varepsilon_{n}|u_{n}(t)\rangle$, with the quasienergy operator
$Q(t)\equiv{H}(t)-\ui\partial_t$.
One can treat the time $t$ as a coordinate under periodic boundary conditions, and choose a complete set of bases $e^{\ui m\omega t}$ labeled by the integer $m$.
In the extended Hilbert space ${\cal F}={\cal H}\otimes {\cal L}_T$, where ${\cal L}_T$ is the space of $T$-periodic functions
and ${\cal H}$ is the Hilbert space, the operator $Q$ can be written as a block-tridiagonal matrix (for harmonic driving), namely,
\begin{align}\label{Qmatrix}
Q=
\left(\begin{array}{ccccc}
 \ddots& & & & \\
V &{H}_{\rm s}+ \omega& V & & \\
 &V &{H}_{\rm s}& V & \\
 & & V & {H}_{\rm s}- \omega& V  \\
& & &  & \ddots
\end{array} \right),
\end{align}
with each block having the same dimension as $H_{\rm s}$. The diagonal blocks $H_{\rm s}+m\omega$ are copies of the static Hamiltonian with energy shifts $m\omega$ ($m=\pm1,\pm2,\cdots$),
and the off-diagonal blocks $V$ couple the two neighboring copies. Copying and shifting unperturbed bands lead to new band crossings (see Fig.~1 in the main text).
The driving $V$ then opens the gaps and brings nontrivial topology.

From Eq.~(\ref{Qmatrix}), one sees that the Floquet system can be treated as a multiband system.
One can find that the lower $n_d/2$ {\it decoupled} $h_0$ bands of $H_{\rm s}$ (with the dispersion $-h_0({\bm k})$) can have crossings with the upper $n_d/2$ $h_0$ bands ($h_0({\bm k})+2n\omega$)  in the blocks $H_{\rm s}+2n\omega$ ($n=0,\pm1,\cdots$), which defines the 0-BISs 
(corresponding to the momenta where $h_{0}({\bm k})=n\omega$), and can also have crossings with the upper $n_d/2$ $h_0$ bands  ($h_0({\bm k})+(2n-1)\omega$) in the blocks $H_{\rm s}+(2n-1)\omega$, which defines the $\pi$-BISs (the momenta where $h_{0}({\bm k})=n\omega+\omega/2$).
After gap openings via ${\bm h}_{\mathrm{so}}$ and $V$, the topology of the lower $n_d/2$ Floquet bands, characterized by $W$, should be contributed from all the $0$- and $\pi$-BISs.
We further define $\nu_j$ as the topological invariant associated with the $j$-th BIS.
Note that the {\it lower} bands for $0$-BISs are the {\it upper} bands for $\pi$-BISs. Therefore, for a $0$-BIS, the contribution to $W$ is $\nu_j$, while for a $\pi$-BIS, the contribution should be $-\nu_j$.
We then have the generalized bulk-surface duality for Floquet systems:
\begin{align}\label{W_nu_S}
W=&\sum_j (-1)^{q_j/\pi}\nu_{j},
\end{align}
where $q_j=0$ or $\pi$ for the $q_j$-BIS.

\subsection{Dynamical characterization}~\label{Sec1B}

In our dynamical characterization scheme,
we examine quench-induced spin dynamics evolving under $H(t)$, and employ the stroboscopic time-averaged spin textures, which read
\begin{align}~\label{yi_S}
\overline{\langle\gamma_i({\bm k})\rangle}=\lim_{N\to\infty}\frac{1}{N}\sum_{n=0}^N {\rm Tr}\big[\rho_0({\bm k}) U^\dagger({\bm k},nT)\gamma_iU({\bm k},nT)\big],
\end{align}
where $\rho_0$ is the density matrix of the initial state.


We first set the driving phase $\phi=0$ in $V(t)=Ve^{\ui(\omega t-\phi)}+V e^{-\ui(\omega t-\phi)}$.
Suppose that the initial state is fully polarized along the $\gamma_0$ axis (BISs are defined by $h_{F,0}({\bm k})=0$) such that $\gamma_0\rho_0=-\rho_0$.
We then have $U(nT)=\exp\big(-\ui H_F\cdot nT\big)=\cos(E_F\cdot nT)-\ui\sin(E_F\cdot nT)H_F/E_F$, with $E_F=\sqrt{\sum_{i=0}^d h_{F,i}^{\,2}}$ being band energy, which leads to
\begin{align}\label{gammai_0}
\overline{\langle\gamma_i({\bm k})\rangle}=\frac{h_{F,i}{\rm Tr}\left[\rho_0H_F\right]}{E_F^2}=-\frac{h_{F,i}({\bm k})h_{F,0}({\bm k})}{E_F^2({\bm k})}.
\end{align}
This result directly yields the dynamical characterization of BISs: $\overline{\langle\gamma_i({\bm k})\rangle}=0$ for all $i$. We note that on the BIS,
\begin{align}
\partial_{k_\perp}\overline{\langle\gamma_i({\bm k})\rangle}=-\lim_{k_{\perp}\to0}\frac{1}{2k_{\perp}}\frac{h_{F,i}+\mathcal{O}(k_{\perp})}
{E_F^{2}+\mathcal{O}(k_{\perp})}\cdot 2k_{\perp}=-\frac{h_{F,i}}{E_F^{2}},
\end{align}
Hence, the dynamical spin-texture field ${\bm g}({\bm k})=(g_1,g_2,\cdots,g_d)$ with $g_i({\bm k})\equiv-\partial_{k_\perp}\overline{\langle\gamma_i({\bm k})\rangle}/{\cal N}_k$ satisfies
${\bm g}({\bm k})\vert_{{\bm k}\in{\rm BIS}}\simeq {\bm h}_{\rm so}^F({\bm k})$.
The topological invariants on BISs can be characterized by the winding of the dynamical field ${\bm g}({\bm k})$, given by Eq.~(5) in the main text.

When the driving phase $\phi$ is nonzero, we first introduce the time-evolution operator from time $t_i$ to $t$, which reads
\begin{align}
U(t,t_i;\phi)={\cal T}\exp\left[-\ui\int_{t_i}^t H(\tau;\phi)d\tau\right].
\end{align}
Here ${\cal T}$ denotes time ordering. We then have
$U(T,0;\phi)=U(T-t_\phi,-t_\phi;0)=U^\dagger(0,-t_\phi;0)U(T,0;0)U(0,-t_\phi;0)$ with $t_\phi\equiv\phi/\omega$,
which yields
\begin{align}~\label{HF_phi_S}
H_F^\phi=U^\dagger(t_\phi)H_FU(t_\phi),
\end{align}
where $U(t_{\phi})\equiv{\cal T}\exp\big[-\ui\int_{0}^{t_{\phi}} H(\tau;\phi=0)d\tau\big]$. Thus, $H_F^{\phi}$ is related to $H_F$ via
a local unitary transformation; they have the same energy spectrum and the same topology.
The dynamical characterization becomes slightly different in even and odd dimensions:
(i) In even dimensions, the dynamical scheme can be directly applied. Since $H_F^{\phi}={\bm h}_F^\phi({\bm k})\cdot{\bm\gamma}=\sum_{i=0}^d h^\phi_{F,i}({\bm k})\gamma_i$,
the BIS is now defined as momenta where $h^\phi_{F,0}({\bm k})=0$, and the spin-orbit field is ${\bm h}_{\rm so}^F=(h^\phi_{F,1},h^\phi_{F,2},\cdots,h^\phi_{F,d})$.
It can be checked that the stroboscopic time-averaged spin textures defined in Eq.~(\ref{yi_S}) yield
$\overline{\langle\gamma_i({\bm k})\rangle}=-h^\phi_{F,i}({\bm k})h^\phi_{F,0}({\bm k})/E_F^2({\bm k})$ when the initial state $\rho_0$ is fully polarized.
(ii) In odd dimensions, the driving will induce extra spin axes in the Floquet Hamiltonian, 
and a direct characterization will fail. We modify the stroboscopic time average by starting from the initial time $t_\phi$,
namely,
\begin{align}~\label{yi_mod_S}
\overline{\langle\gamma_i\rangle}=\lim_{N\to\infty}\frac{1}{N}\sum_{n=0}^N {\rm Tr}\big[\rho_0(t_\phi)U^\dagger(nT+t_\phi,t_\phi;\phi)\gamma_iU(nT+t_\phi,t_\phi;\phi)\big],
\end{align}
where $\rho_0(t_\phi)=U(t_\phi)\rho_0 U^\dagger(t_\phi)$.
We note that $U(nT+t_\phi,t_\phi;\phi)=U(nT,0;0)=\exp(-\ui H_F\cdot nT)$. For such an incompletely polarized initial state $\rho_0(t_\phi)$,
spin textures in Eq.~(\ref{yi_mod_S}) can still be used to characterize the topology of $H_F$~\cite{Zhanglong2019b_S}.  In fact, we have
$\overline{\langle\gamma_i\rangle}=-h_{F,i}h^\phi_{F,0}/E_F^2$. Topological patterns on the BISs $h^\phi_{F,0}({\bf k})=0$ characterize the winding of ${\bm h}^{F}_{\rm so}({\bm k})$.

\subsection{General driving fields}~\label{Sec1C}

In even dimensions, our dynamical characterization scheme is generally valid for an arbitrary driving field $V(t)=V(t+T)$, 
as long as the BISs are well-defined.
(One exception is that the nondriven part $H_{\rm s}$ has only constant bands, e.g., the stepwisely driven model introduced in Ref.~\cite{Kitagawa2010_S}.)
A periodic driving field can be generally written as $V(t)=\sum_{n>0}V^{(n)} e^{\ui n\omega t}+V^{(n)\dagger} e^{-\ui n\omega t}$, with the corresponding quasienergy operator taking the form
\begin{align}\label{Qmatrix_General}
Q=
\left(\begin{array}{ccccccc}
\ddots &\vdots &\vdots &\vdots & \vdots&\vdots & \adots \\
\cdots &{H}_{\rm s}+ 2\omega &V^{(1)} & V^{(2)} & V^{(3)} &V^{(4)} &\cdots \\
\cdots &V^{{(1)}\dagger} &{H}_{\rm s}+ \omega& V^{(1)} & V^{(2)} &V^{(3)} &\cdots \\
\cdots&V^{{(2)}\dagger} &V^{{(1)}\dagger} &{H}_{\rm s}& V^{(1)} & V^{(2)}&\cdots\\
\cdots& V^{{(3)}\dagger} & V^{{(2)}\dagger} & V^{(1)\dagger} & {H}_{\rm s}- \omega& V^{(1)}& \cdots \\
\cdots& V^{{(4)}\dagger} & V^{{(3)}\dagger} & V^{(2)\dagger} &  V^{(1)\dagger}  & {H}_{\rm s}- 2\omega& \cdots \\
\adots &\vdots &\vdots &\vdots & \vdots & \vdots& \ddots
\end{array} \right).
\end{align}
One can see that compared to the case of harmonic driving ($V^{(n)}=0$ for all $n>1$), the BISs for general driving fields
are also formed by the crossings of the $h_0$ bands in the diagonal blocks ${H}_{\rm s}+m\omega$,
while the higher frequency terms $V^{(n>1)}$ only provide additional couplings to open the gaps.
Therefore, we can make a similar argument to the one for deriving Eq.~(\ref{W_nu_S}), and conclude that the generalized bulk-surface duality should hold for general driving fields.

In odd dimensions, the periodic driving necessitates a constraint $V(t_\phi+t)=V(t_\phi-t)$ with $0\leq t_\phi<T$, which ensures a chiral symmetry inherited from the static Hamiltonian $H_{\rm s}$. 
We first deal with the special case $t_\phi=0$, i.e., $V(t)=V(-t)$. In this case, we have $S H({\bm k},t)S^{-1}=-H({\bm k},-t)$, where $S$ denotes the chiral symmetry of the static Hamiltonian.
We further have $S U({\bm k},t)S^{-1}=U({\bm k},-t)$, which yields $S H_F({\bm k})S^{-1}=-H_F({\bm k})$.
Thus, the Floquet Hamiltonian also has the chiral symmetry $S$, and can be written as
\begin{align}~\label{HF_S}
H_F({\bm k})=h_{F,0}({\bm k})\gamma_0+\sum_{i=1}^d h_{F,i}({\bm k})\gamma_i,
\end{align}
where the $\gamma$ matrices are those in $H_{\rm s}$.
The topology of $H_F$ then can be detected via the measurement of stroboscopic time-averaged spin textures $\overline{\langle\gamma_i({\bm k})\rangle}$ defined in Eq.~(\ref{yi_S}).
For a general case with $t_{\rm \phi}\neq0$, we suppose that $t_\phi$ can be controlled by a tunable phase $\phi$ via  $t_\phi\equiv(\phi/2\pi)T$, which yields the relation (\ref{HF_phi_S}). 
It is obvious that $H_F^{\phi}$ has a different chiral symmetry $S_{\phi}=U^\dagger(t_\phi)SU(t_\phi)$, and thus {\it cannot} be written in the form of Eq.~(\ref{HF_S}). 
But the dynamical characterization can be accomplished via the modified time averages defined in Eq.~(\ref{yi_mod_S}).

\begin{figure}
\includegraphics[width=0.9\textwidth]{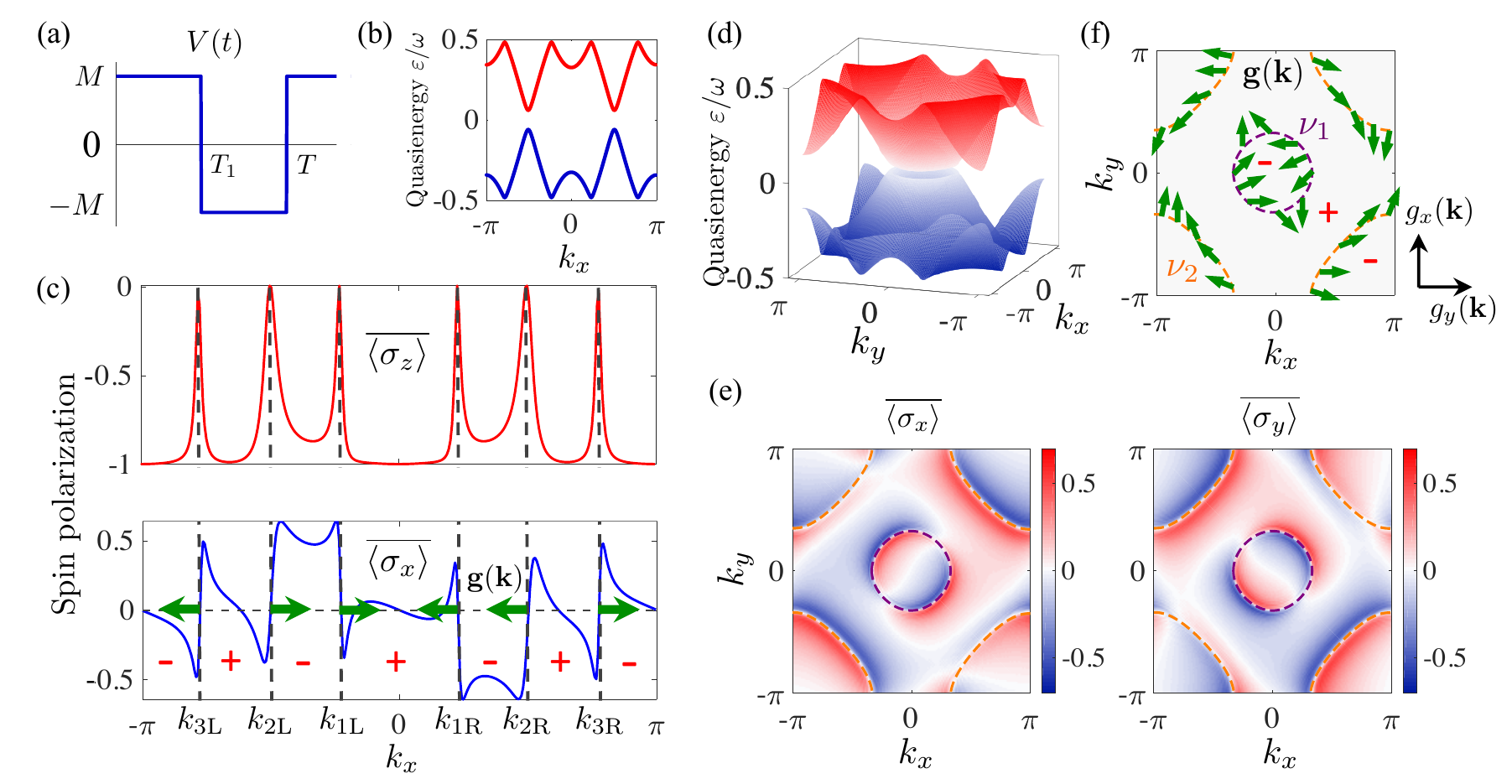}
\caption{Dynamical characterization for square-wave driving. (a) The applied square-driving field $V(t)=M(t)\sigma_z$, where $M$ and $T_1$ are both tunable.
(b) The Floquet bands of the 1D driven model with $M=10t_0$, $T_1=0.5T$ and $\omega=3t_0$. (c) Dynamical characterization of the 1D model in (b)
via stroboscopic time-averaged spin textures $\overline{\langle\sigma_{x,z}(k_x)\rangle}$. The stroboscopic time averages are taken from the initial time $t_i=T_1/2$.
The constructed dynamical field ${\bm g}(k_x)$ characterizes the topological invariants $\nu_{1,2}=-1$ and $\nu_{3}=+1$, which gives $W_0=\nu_2=-1$ and $W_\pi=\nu_1+\nu_3=0$.
(d) The Floquet bands of the 2D driven model with $M=15t_0$, $T_1=0.6T$ and $\omega=8t_0$. 
(e) Stroboscopic time-averaged spin textures $\overline{\langle\sigma_{x,y}({\bm k})\rangle}$, where two BISs are identified.
(f) The dynamical field ${\bm g}({\bm k})$ constructed by the spin textures in (e) characterizes the topological invariants $\nu_1=-1$ and $\nu_2=+1$, giving the Chern number ${\rm Ch}=-2$. 
In both 1D and 2D cases, we set $t_{\rm so}=t_0$ and $m_z=0$.
}\label{figS1}
\end{figure}

Here we take the square-wave driving $V(t)=M(t)\sigma_z$ as an illustration [see Fig.~\ref{figS1}(a)], where
\begin{align}
M(t)=\left\{ \begin{array}{lr}
M &  0\leq t<T_1\\
-M &  T_1\leq t<T,
\end{array} \right.
\end{align}
which is a form of polychromatic driving and often applied in real experiments.
We still drive the 1D chiral topological phase and 2D quantum anomalous Hall model as in the case of harmonic driving.
The numerical results for these two models are shown in Fig.~\ref{figS1}(b-c) and (d-f), respectively.
 In (c), the stroboscopic time averages are taken from the initial time $t_i=T_1/2$ such that $\overline{\langle\sigma_y(k_x)\rangle}=0$.
The dynamical field ${\bm g}(k_x)$ obtained from $\overline{\langle\sigma_x(k_x)\rangle}$ then characterizes the topology of the 1D driven model.
In (e), the stroboscopic time averages are taken directly from $t_i=0$.
Note that the amplitude of the square-wave driving is set to be $M = 10t_0$ (for the 1D model) or $15t_0$ (for the 2D model), much larger than other parameters. 
Therefore, the results also show clearly that our dynamical characterization scheme is applicable to strong drives.

\begin{figure}
\includegraphics[width=0.7\textwidth]{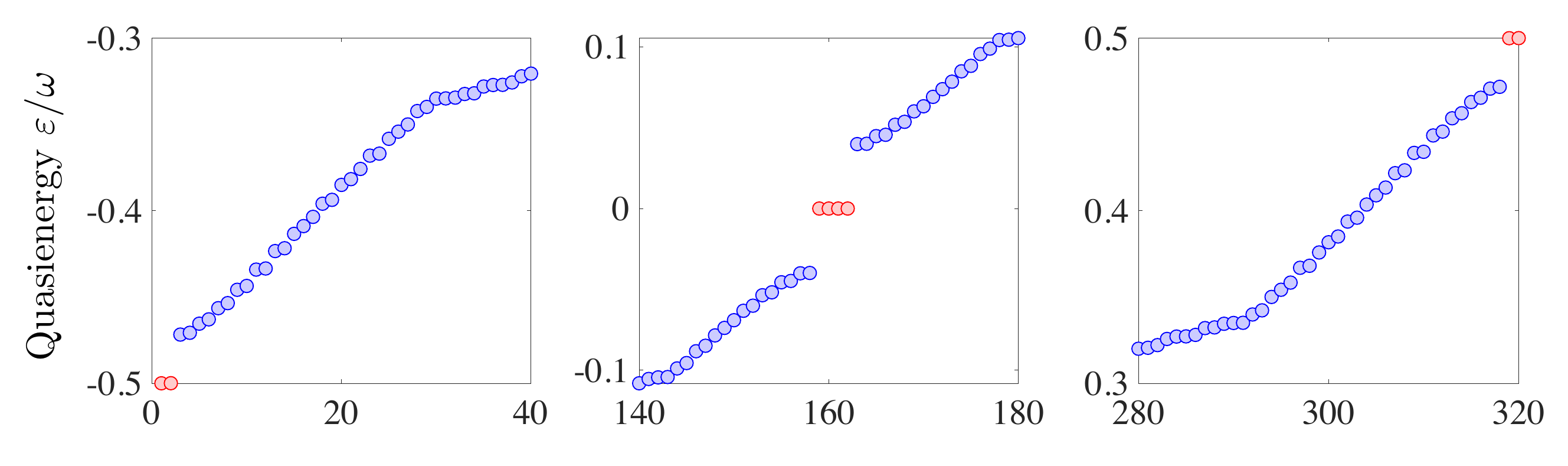}
\caption{Quasienergy spectrum of the 1D harmonically driven model under open boundary conditions. The number of edge states (red circles) at the gap around $\varepsilon=0$ (or $\varepsilon=\pi/T$)
equals to the topological invariants on $0$-BISs (or $\pi$-BISs).
The parameters are taken the same as in Fig.~1 of the main text, with the lattice site number set to be 160.
}\label{figS2}
\end{figure}

\section{Relation with the number of boundary states}~\label{Sec2}

In our dynamical characterization, the topological invariants on BISs directly reflect the number of boundary states at the two gaps $\varepsilon=q/T$ ($q=0,\pi$), i.e.,
\begin{align}
n_{\rm edge}(\varepsilon)=\sum_{j\in\,q\textrm{-BIS}} v_j=W_q
\end{align}
For the 1D periodically driven model considered in the main text, here we give the quasienergy spectrum under open boundary conditions (see Fig.~\ref{figS2}).
We take the same parameters as in Fig.~1, and find that the number of edge states at the two gaps are both 2, being identical to the topological invariants $W_{0,\pi}$.
The verification for 2D systems has been shown in Fig.~2 of the main text.

\section{Identifying the emergence of topological singularities in phase bands}~\label{Sec3}

\begin{figure}
\includegraphics[width=0.9\textwidth]{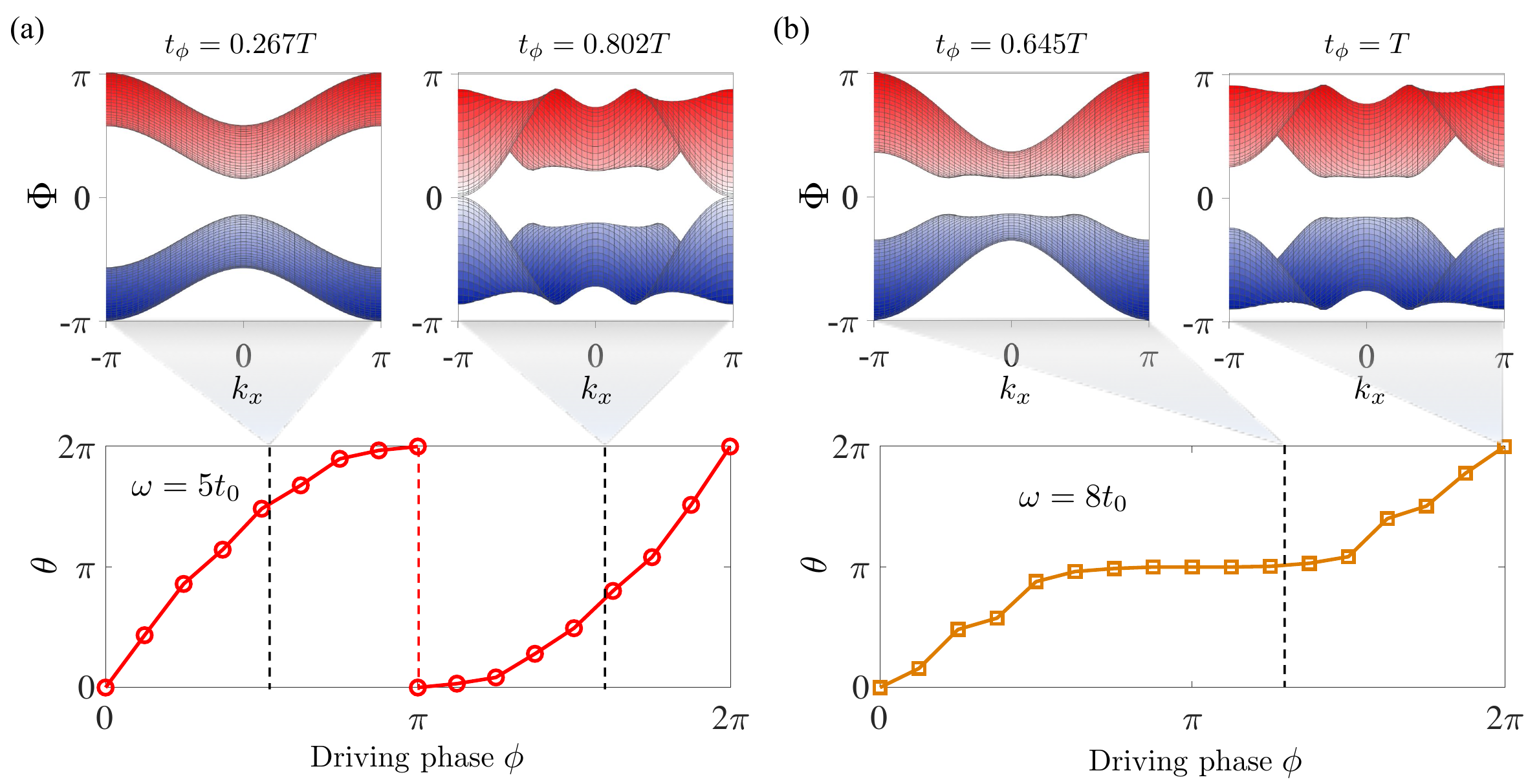}
\caption{Rotation angle $\theta$ serves as a signal of the emergence of topological singularities in the phase bands of the evolution operator $U(t_\phi)$.
Upper panel: Phase bands of $U(t_\phi)$. Lower panel: the measured rotation angle $\theta$ as a function of the driving phase $\phi$.
Two cases are considered: $\omega=5t_0$ (a) and  $\omega=8t_0$ (b).
In (a), two driving-induced topological singularities emerge at $t_\phi=0.267T$ and $t_\phi=0.802T$, respectively.
In (b), only one driving-induced singularity emerge ($t_\phi=0.645T$), which corresponds to the $\pi$-BIS in the Floquet bands ($t_\phi=T$).
The emergence of these singularities is captured by the evolution of the rotation angle.
}\label{figS3}
\end{figure}

The evolution operator  $U({\bm k},t)$ can be diagonalized as
\begin{align}~\label{Ut_S}
U({\bm k},t)=e^{-\ui\Phi({\bm k},t)}|+\rangle\langle+|+e^{\ui\Phi({\bm k},t)}|-\rangle\langle-|,
\end{align}
where $\pm\Phi({\bm k},t)$ ($0\leq \Phi\leq \pi$) are called phase bands~\cite{Nathan2015_S}.
It has been shown that topologically-protected crossings of the phase bands throughout the driving period $0<t<T$ determines the topology of Floquet systems~\cite{Nathan2015_S}.
For our 2D model with the setting $m_z>0$, besides the gap-closing stemming from the topology of static bands, {\it driving}-induced gap-closings all occur at the $M$ point (see Fig.~\ref{figS3})
(When $m_z<0$, the situation is reversed: ``static'' gap-closing at $M$ and driving-induced gap-closing at $\Gamma$).
We consider the neighboring region of the $M$ point, where the evolution operator can be approximated as $U(t)\approx \exp[-\ui\Phi(t)\sigma_z]$.
Note that for the vector field ${\bm h}_F({\bm k})$ on a BIS surrounding the $M$ point ($h_{F,z}=0$), the transformation (\ref{HF_phi_S}) can be regarded as
a rotation about the $\sigma_z$ axis by an angle $2\Phi(t_\phi)$, which generates the $\phi$-dependent field ${\bm h}^\phi_F({\bm k})$.
Due to the expression (\ref{Ut_S}), a topological singularity emerges when $\Phi(t_\phi)$ reaches $\pi$ or $0$; at the meantime, the field ${\bm h}_F({\bm k})$ on the BIS
should be rotated in a full circle by the operator $U(t_\phi)$.
We can expect that the emergence of topological singularities can be revealed by examining how spin textures evolve with the increasing driving phase $\phi$.

We choose the two poles $h^\phi_{F,y}/|{\bm h}^\phi_{F}|=\pm1$ to detect the rotation, which can be identified by the stroboscopic time-averaged spin texture $\overline{\langle\sigma_{x}({\bm k})\rangle}$. According to Eq.~(\ref{gammai_0}), dynamical characterization of $\overline{\langle\sigma_{x}({\bm k})\rangle}=0$ identifies both the BISs ($h_{F,z}=0$) and
the curves $h_{F,x}({\bm k})=0$. Thus, the poles $h^\phi_{F,y}/|{\bm h}^\phi_{F}|=\pm1$ are the intersections of these two kinds of curves [see Fig.~4(c) in the main text].
We measure the rotation angle $\theta$ and plot it as a function of the driving phase $\phi$. The results are shown in Fig.~\ref{figS3} and also Fig.~4(e) in the main text,
which confirm our analysis. 

The above results can be generalized to 2D systems with general driving fields.
For a driving field $V(t)=V_0(t)\sigma_z$ with $V_0(t)$ taking an arbitrary form, 
one can tune the driving phase $\phi$ in the field $V(t-t_\phi)$ with $t_\phi=(\phi/2\pi)T$.
With this manipulation, we always have the relation (\ref{HF_phi_S}). 
One can easily check that the above argument for identifying the emergence of topological singularities works in this general case.

\section{Experimental realization}~\label{Sec4}

In this section, we present more details for the experimental realization of our dynamical approach.
We first emphasize that both the 1D and 2D periodically driven models proposed in the main text are readily accessible: 
Their static parts $H_{\rm s}$  have been realized in ultracold atom~\cite{Song2018_S,Wu2016_S,Sun2017_S,Sun2018_S,Yi2019_S} and solid state~\cite{WangY2019_S} systems;
the harmonic (or square-wave, see Sec.~\ref{Sec1C}) driving field can be easily implemented by modulating the bias magnetic field.
Besides, our dynamical scheme employs the quench dynamics of a polarized trivial state, which, compared to the eigenstates of the Floquet bands, is easy to prepare in experiment.

The observation of the generalized dynamical bulk-surface correspondence can basically follow that carried out in nondriven systems~\cite{Sun2018_S,Yi2019_S,WangY2019_S,Song2019_S,Niu2020_S,Xin2020_S,Ji2020_S,Wang2020_S}. 
Three points should be noted when performing the measurement: 
(i) It can be proved that measuring the $i$th spin component in quenching $j$th spin axis precisely equals to measuring the $j$th spin component in quenching the $i$th spin axis~\cite{Zhanglong2019b_S}.
Hence, for cold atom experiments, the stroboscopic time-averaged spin textures in all directions can be obtained by 
measuring only one single spin component (e.g., the $\sigma_z$-component) after a series of quenches along different spin quantization axes~\cite{Yi2019_S}.
(ii) Although the stroboscopic time averages are defined over an infinite time interval [see Eq.~(3) in the main text],
short-term quench dynamics already provides enough information to characterize the topology.
In the experiment of dynamically characterizing 2D quantum anomalous Hall model~\cite{Yi2019_S},
the time averages are taken over only two or three oscillation periods.
Hence, in experiment on Floquet topological phases, one can take stroboscopic time averages over a finite interval of several $T_\Delta$. Here $T_\Delta=2\pi/\Delta$
denotes the period of the low-frequency oscillation, with $\Delta$ being the local energy gap of the Floquet bands. 
For example, $T_\Delta\sim 20T$ in Fig.~3(a) in the main text.
This makes the dynamical characterization robust against heating effects.
(iii) In odd dimensions, the modified initial time $t_i=t_\phi$ can be determined by
measuring the spin dynamics in the axis of the chiral symmetry $S$ of the static Hamiltonian (e.g., the $\sigma_y$ axis for the proposed 1D model) and averaging it from different initial times; 
when the stroboscopic time average starts from $t_i=t_\phi$, we should have $\overline{\langle S\rangle}=0$ for all ${\bm k}$.



\end{document}